\RequirePackage{fix-cm}
\documentclass[smallextended]{svjour3}       
\smartqed  
\usepackage{braket}
\usepackage{pdfpages}
\usepackage{pgfplots}
\usepackage{csquotes}
\usepackage{float}
\usepackage{multirow}
\usepackage{dcolumn}
\usepackage[colorlinks=true, citecolor=blue]{hyperref}
\usepackage{hhline}
\usepackage{amssymb}
\usepackage{amsmath}
\begin{document}

\title{Solving Diner's Dilemma Game, Circuit Implementation and Verification on the IBM Quantum Simulator}

\author{Amit Anand \and Bikash K. Behera $^*$ \and Prasanta K. Panigrahi}

\institute{Amit Anand \at Department of Mechanical Engineering, Indian Institute Of Engineering Science And Technology, Shibpur, Howrah-711103, West Bengal, India\\ \email{amitanand844@gmail.com} \and Bikash K. Behera \at Bikash's Quantum (OPC) Pvt. Ltd., Balindi, Mohanpur, 741246, Nadia, West Bengal, India\ \at Department of Physical Sciences, Indian Institute of Science Education and Research Kolkata, Mohanpur 741246, West Bengal, India\\ \email{bikash@bikashsquantum.com},\ $^{*}$Corresponding author \and Prasanta K. Panigrahi \at Department of Physical Sciences, Indian Institute of Science Education and Research Kolkata, Mohanpur 741246, West Bengal, India \\ \email{pprasanta@iiserkol.ac.in}}

\date{Received: date / Accepted: date}

\maketitle

\begin{abstract}
Diner's dilemma is a problem of interest to both economics and game theory. Here, we solve this problem for n = 4 (where n is the number of players)with quantum rules.  We are able to remove the dilemma of diners between the Pareto optimal and Nash equilibrium points of the game. We find the quantum strategy that gives maximum payoff for each diner without affecting the payoff and strategy of others. Quantum superposition and entanglement is used as a resource which gave the supremacy over any classical strategies. We present the circuit implementation for the game, design it on the IBM quantum simulator and verify the strategies in the quantum model.
\end{abstract}


\section{Introduction}\label{qdd_Sec1}
Game theory is the science of strategy of optimal decision-making \cite{qdd_guo}, or of independent and competing players in a strategic setting \cite{qdd_Gametheorybook},\cite{qdd_luca,qdd_iqbal,qdd_kolokoltsov}. It provides a framework based on the construction of mathematically rigorous models that describe situations of conflict and cooperation between rational decision-makers \cite{qdd_khangaming}. In decision theory and economics, rational behaviour is defined as choosing actions that maximize one's Payoff (or some form of Payoff) subject to constraints that one faces \cite{qdd_BrunnerLinden2013}. Game theory has been successfully applied to many relevant situations, such as business competition, the functioning of the market, political campaigning, jury voting, auctions, many more \cite{qdd_manik2016}. It is also used in the field of evolutionary biology and psychology. John Von Neumann's paper ``On the Theory of Games of Strategy" \cite{qdd_johnvonneumann1928} in 1928 led the foundation of modern game theory. His paper was followed by his 1944 book ``Theory of Games" and ``Economic Behavior" co-authored with Oskar Morgenstern \cite{qdd_johnandmorgenstern1944}. The second edition of this book provided an axiomatic theory of utility, which reincarnated Daniel Bernoulli's old theory of utility (of the money) as an independent discipline. Von Neumann's work in game theory culminated in this 1944 book. Till that time, significant discussions were on cooperative games. In 1950, the first mathematical discussion of the prisoner's dilemma appeared. Around this same time, John Nash developed a criterion for mutual consistency of players' strategies, known as \textit{Nash equilibrium} \cite{qdd_nashquilibrium1950,qdd_iqbalequi}, applicable to a wider variety of games than the criterion proposed by Von Neumann and Morgenstern.

Nash proved that every finite n-player, non-zero-sum (not just 2-player zero-sum) and non-cooperative game has a Nash equilibrium in mixed strategies. The promise of a Nash equilibrium solution is a foundational concept for game theory as it may be used to guarantee the behaviour for the non-cooperating players. In conventional games, the relative simplicity of the proof of Nash's theorem for the existence of an equilibrium in mixed strategies entirely relies on Kakutani's fixed-point theorem \cite{qdd_kakutani}. For quantum games, Meyer \cite{qdd_meyer} established the existence of Nash equilibrium in mixed strategies, which are modelled as mixed quantum states, using Glicksberg's \cite{qdd_glicksberg} extension of Kakutani's fixed-point theorem to topological vector spaces. Khan and Humble \cite{qdd_khanembedding} show in their work that the Kakutani fixed-point theorem does not apply directly to quantum games played with pure quantum strategies. But, one can use Nash's embedding of compact Riemannian manifolds into Euclidean space \cite{qdd_nashembedding,qdd_khanreview} (Nash's other, mathematically more famous theorem) and, under appropriate conditions indirectly apply the Kakutani fixed-point theorem to guarantee Nash equilibrium in pure quantum strategies.

In our work, we solved a well-known problem in game theory and economic theory, i.e. Diner's Dilemma in the quantum domain. Game theory is an essential discipline of Applied Mathematics which has many applications in economics, psychology, and biology that are probabilistic in nature to a great extent. This is the main reason for quantizing this game. We use two main features of quantum physics, such as entanglement \cite{qdd_du} and non-locality. By using the non-local correlations, we gain an advantage over classical correlations. Since the game is non-cooperative (i.e. participants cannot interact with each other once the game has started), entanglement plays a vital role in deciding their strategy. Eisert et al. \cite{qdd_ewl} showed that their quantum computational implementation of Prisoner's Dilemma produced non-classical correlations and resolved the dilemma (Nash equilibrium is also optimal). They have introduced an equivalence principle which guarantees that the performance of a classical game and its quantum extension can be compared in an unbiased manner. In Ref. \cite{qdd_shimamura}, Shimamura et al. establish a more robust result that entanglement enabled correlations always resolve dilemmas in nonzero-sum games, and that classical correlations do not necessarily do the same. Quantum entanglement is clearly a resource for quantum games.

In 2004, Gneezy, Haruvy and Yafe \cite{qdd_splittingthebill2004} did a social experiment entitled ``The Inefficiency Of Splitting The Bill", in which six individuals were made to dine together in a restaurant. In that social experiment, they test the hypothesis based on standard economic assumptions that consumers will find it optimal to increase consumption when marginal benefit exceeds marginal cost and to lower consumption when the opposite holds. The six participants were not allowed to communicate among themselves or waiters once the game start and they can place there order by writing it on paper. They produce their results for four different cases: (1) each participant pays their bill individually, (2) bill was equally split among six participants, (3) bill was paid by restaurant owner, (4) participants paid only 1/6 of his/her bill and rest was paid by the restaurant owner. Fig. \ref{qdd_Fig1} below shows the results of the first three of the four cases. They found that the bill was more when they were splitting the bill evenly than when they are paying individually. The efficiency implication of the different payment methods is straight forward. When splitting the bill, diners consume such that the marginal social cost they impose is larger than their own marginal utility and, as a result, they over-consume relative the social optimum. This makes case-2 very interesting.  In fact, it is easy to show that in a classical setting the only efficient payment rule is the individual one. It turns out that subjects’ preferences are consistent with increasing efﬁciency. When asked to choose, prior to ordering, whether to split the bill or pay individually, 80\% choose the latter. That is, they prefer the environment without externalities. However, in the presence of externalities, they nevertheless take advantage of others. 

\begin{figure}[]
\centering
\includegraphics[width=12cm]{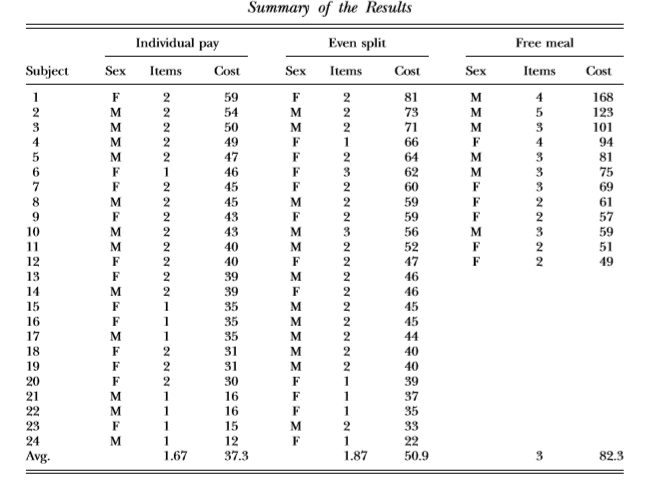}
\caption{Results of first three cases of the social experiment entitled "The Inefficiency Of Splitting The Bill"}
\label{qdd_Fig1}
\end{figure}

Here, we solve only for the case-(2) with four participants. This is the most interesting case in which participants face a dilemma in deciding their strategy while placing the order. They do not want to increase their marginal cost by increasing the consumption, but they also do not want to lower their marginal benefits. The setting of a game is such that they can either order cheap food (denoted by C) or expensive food (denoted by E). Each participant is unaware of the order placed by the others. They cannot make any strategy for placing order depending on the strategy others are taking. According to the strategy taken, each player will then be awarded the payoff value. Each player aims to increase their individual Payoff.

IBM Q gives access to a superconducting-qubit based operating system that is globally access to a wide class of researchers and has found significant applications in a user-friendly interface \cite{qdd_IBM}. A number of experiments in the field of quantum simulations \cite{qdd_AggarwalarXiv2018,qdd_ZhukovQIP2018,qdd_MalikRG2019,qdd_SchuldEPL2017,qdd_TannuarXiv2018,qdd_ManabputraarXiv2018,qdd_ViyuelanpjQI2018} developing quantum algorithms \cite{qdd_GarciaJAMP2018,qdd_RounakarXiv2018,qdd_GangopadhyayQIP2018,qdd_DeffnerHel2017,qdd_YalcinkayaPRA2017,qdd_SrinivasanarXiv2018,qdd_DasharXiv2018,qdd_Baishya1RG2019,qdd_Baishya2RG2019}, testing of quantum information theoretical tasks \cite{qdd_HuffmanPRA2017,qdd_SwainQIP2019,qdd_AlsinaPRA2016,qdd_KalraarXiv2017,qdd_VishnuQIP2018,qdd_quantumsecretsharing}, quantum cryptography \cite{qdd_BeheraQIP2017,qdd_Plesa2018,qdd_MajumderarXiv2017,qdd_SarkarRG2019}, quantum error correction \cite{qdd_GhoshQIP2018,qdd_Roffe2018,qdd_SatyajitQIP2018,qdd_HarperarXiv2018,qdd_SingharXiv2018}, quantum applications \cite{qdd_DasharXiv2017,qdd_Solano2arXiv2017,qdd_BeheraQIP2019,qdd_BeheraarXiv2018}, quantum games \cite{qdd_PalRG2018,qdd_SinghRG2019}, quantum chemistry \cite{qdd_KumararXiv2019}, quantum teleportation \cite{qdd_SisodiaQIP2017,qdd_RajiuddinRG2019,qdd_ChatterjeeRG2019}, quantum neural network \cite{qdd_DashRG2019}, quantum machine learning \cite{qdd_DuttaarXiv2018,qdd_ShenoyRG2019}, quantum walk \cite{qdd_AdhikariRG2019}, quantum robotics \cite{qdd_MahantiQIP2019,qdd_MishraRG2019} have been performed on the IBM Q experience platform.

We design quantum circuits and simulate them by using IBM quantum experience platform. We use the `IBM Q simulator' for verifying all the rules of the game. Appropriate quantum circuits for the unitary operators are designed, and circuit implementation by the use of single-qubit and two-qubit controlled is appropriately explained. We take four qubits on the IBM Q simulator to design our circuit and perform the experiment. We successfully verify the protocol for diner's dilemma game on the IBM quantum computer.

The paper is organized as follows. In Section \ref{qdd_Sec2}, we solve for the classical model of diner's dilemma game. In Section \ref{qdd_Sec3}, we present the quantum model of the game. Following which, we implement the above game on the IBM quantum computer and present the results in section \ref{qdd_Sec4}. Finally, we conclude in Section \ref{qdd_sec5} and discuss the future directions of this work.

\section{Classical Model \label{qdd_Sec2}}

\begin{figure}[]
\centering
\includegraphics[width=12cm]{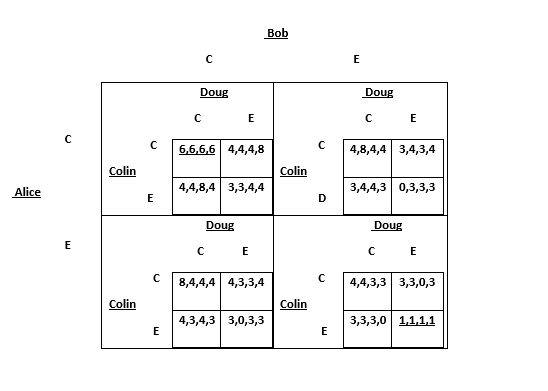}
\caption{Classical payoff table for diner's dilemma for n=4.}
\label{qdd_Fig2}
\end{figure}

The classical diner's dilemma is a non-cooperative, non-strictly competitive, symmetric game. There are four players Alice (A), Bob (B), Colin (C) and Doug (D). Each player has two strategic options, either ordering cheap food (C) or expensive food (E). Depending on the strategies taken, they were assigned a payoff value. Let $\alpha$ represent the joy of eating the expensive meal, $\beta$ the joy of eating the cheap meal, $\gamma$ is the cost of the expensive meal, $\delta$ be the cost of the cheap meal. For assigning the payoff for different cases, we assume $\gamma$-$\delta$ is greater than $\alpha$-$\beta$.  The value of Payoff is decided by the difference between the marginal benefits and marginal cost. If the difference is maximum then they are given payoff value of 8 and when it is minimum then they are given payoff value as 0. For example, when everyone is ordering cheap food or expensive food, there is no difference between marginal benefits and marginal cost but they are given payoff value as 6 in first case and 1 in other due to the first assumption that is $\gamma$-$\delta$ is greater than $\alpha$-$\beta$. When one(let's say A) is ordering the cheap  food and other three are ordering the expensive food, then the difference between marginal benefit and marginal cost is maximum for A. So the payoff assign to A participants is 0 and others were given 3. But when three of them orders cheap food(let's say B, C and D) and one orders expensive food(let's say A), the difference between marginal cost and marginal benefit for A is maximum and gets payoff value as 8 but for the rest three the difference is not minimum and in this case they get a payoff value as 4.The classical payoff value is given in Fig. \ref{qdd_Fig2}. The payoff of Doug can be calculated using,

\begin{eqnarray}
\label{qdd_Eqn1}
Pf_{D}&=&6P(0000)+8P(0001)+4P(0010)+4P(0011)+4P(0100)+4P(0101)\nonumber \\ &+&3P(0110)+3P(0111)+ 4P(1000)+4P(1001)+3P(1010)+3P(1011) \nonumber\\ &+& 3P(1100)+3P(1101)+0P(1110)+1P(1111)
\end{eqnarray}   

where P(wxyz) means probability of selecting strategy w, x, y, z by A, B, C and D respectively and w, x, y, z belongs to either strategy  C or E.  Payoff of Alice, Bob, and Colin can be calculated similarly using classical payoff box. From the payoff Fig. \ref{qdd_Fig2}, it can be seen that self-serving people will choose the strategy E and thus it's Nash equilibrium (NE) point.

\textit{Nash equilibrium} is a play of \textit{T} in which every player employs a strategy that is the best reply, with respect to his preferences over the outcomes, to the strategic choice of every other player. In other words, unilateral deviation from a Nash equilibrium by any one player in the form of a different choice of strategy will produce an outcome which is less preferred by that player than before. Following Nash, we say that a play $ P^{'}$ of \textit{T}  counters another play P if $$ \textit{T}i(P) \leq \textit{T}i(P^{'}) $$ for all players i, against the (n - 1) strategies of the other players in the countered n-tuple, and that a self-countering play is a Nash equilibrium. For Alice, whatever strategy taken by the other three participants, her best reply is E. Since the game is symmetric in nature, it is same for the other three participants, and strategy E is called a dominant strategy. Therefore the strategy taken by ABCD is EEEE (1111) since its Nash Equilibrium point, which gives a payoff of 1 to each player. There is a point in a payoff table which gives maximum payoff to each individual without decreasing the payoff of others, i.e. CCCC (0000) which gives the payoff of 6 to each player. Point (0000) in the payoff table is known as \textit{Pareto Optimal} point. This creates a dilemma in the players' mind. They can only achieve  maximum payoff by mutual cooperation which is not allowed in this setting of a game.

\section{Quantum Model}\label{qdd_Sec3}
In the quantum model, we used  the EWL protocol \cite{qdd_ewl} for quantizing the game. We assign the two basis vectors $\ket{C}=\left [{\begin{array}{c} 1 \\0 \end{array}}\right ]$ and $\ket{E}=\left [{\begin{array}{c} 0 \\1 \end{array}}\right ]$ in the Hilbert space of a two-level system i.e., qubit, the two possible outcomes of classical strategy C and E (cheap food and expensive food respectively). At any point, state of the game is described by a vector in the tensor product space which is spanned by the classical game basis $\ket{WXYZ}$ where (W,X,Y,Z) $\in$(0,1). Four states $\ket{C}$, $\ket{C}$, $\ket{C}$ and $\ket{C}$ are produced by identical sources. The initial state of the game is described by $\ket{\psi_0}=\ket{CCCC}$, where the first qubit is with Alice, second with Bob, third with Colin and fourth with Doug. An operator,

\begin{equation}
\label{qdd_Eq2}
\hat{J}=\frac{1}{\sqrt{2}}(\hat{}{I}\otimes\hat{I}\otimes\hat{I}\otimes\hat{I} + i(i\sigma_y)\otimes (i\sigma_y)\otimes(i\sigma_y)\otimes(i \sigma_y)
\end{equation}
is defined to create an entanglement, where $\hat{I}$ is an Identity and $\sigma_y$ is Pauli-Y operator. On applying the operator given in Eq. \eqref{qdd_Eq2} on $\ket{\psi_0}$, we get a final state as $\ket{\psi_i}$ where

\begin{equation} 
\label{qdd_Eq3}
\ket{\psi_i}=\frac{1}{\sqrt{2}}\ket{0000} + i\ket{1111}
\end{equation}
Now players introduce their quantum strategies
\begin{equation}
\label{qdd_Eq4}
\hat{U}(\theta_K,\phi_K)=\left [{\begin{array}{cc} 
          \\
          e^{i\Phi_k}cos\theta_k/2  &  sin\theta_k/2 \\
          -sin\theta_k/2          &  e^{-i\Phi}cos\theta_k/2
        \end{array}}
        \right ]
\end{equation} 
where $\theta_k\in[0,\pi]$,$\phi_k\in[0,\pi/2]$ and $k\in(A,B,C,D)$. Players then apply their respective operators (or strategies) i.e., $\hat{U_A}$, $\hat{U_B}$, $\hat{U_C}$ and $\hat{U_D}$ (Eq. \eqref{qdd_Eq4}) on $\hat{J}\ket{\psi_0}$. At the end, we use disentangling operator $\hat{J}^\dagger$, the state becomes to $\ket{\psi_f}$. 
\begin{equation}
\label{qdd_Eq5}
\ket{\psi_f}=\hat{J}^\dagger\hat{U_A}\otimes\hat{U_B}\otimes\hat{U_C}\otimes\hat{U_D} \hat{J}\ket{\psi_0} 
\end{equation}
\begin{figure}[]
\centering
\includegraphics[width=12cm]{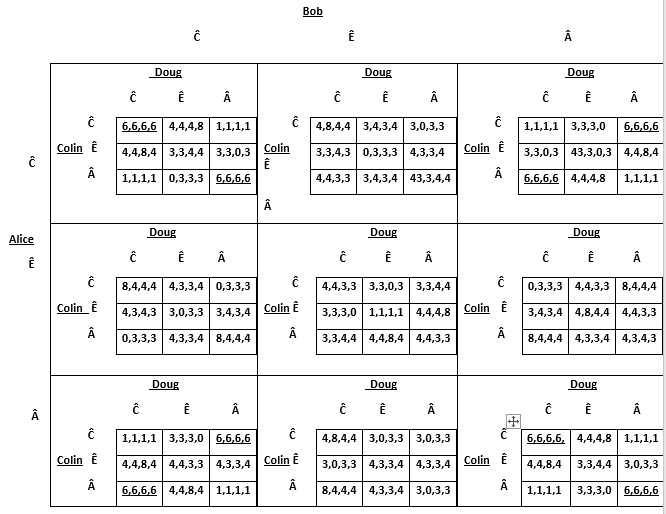}
\caption{Quantum payoff table for diner's dilemma for n=4.}
\label{qdd_Fig3}
\end{figure}

Payoff of Doug can be calculated by using Eq. \eqref{qdd_Eqn1}, where $P(X_A,X_B,X_C,X_D)$ is the joint probability that final state of qubits with the players will collapse to $X_A,X_B,X_C,X_D\in(C,E)$ on measuring $P=|\braket{X_A,X_B,X_C,X_D|\psi_f}{}|^2$ using Eq. \eqref{qdd_Eq5}. Here in our game, we have set of strategies for the entangled states whose has no counterparts in classical domain. If all the players choose to play with $\theta=0$ and $\phi=0$ then the game reduces to local correlations and shows local correlations. However, it shows non local correlations if $\phi\neq 0$. We define three operators or quantum strategy $\hat{C}$, $\hat{E}$, and $\hat{A}$ where,

\begin{equation}
\label{qdd_Eq6}
\hat{C}= \hat{U}(0,0)=\left [{\begin{array}{cc} 
          \\
          1  &  0 \\
          0  &  1
        \end{array}}
        \right ]
\end{equation}
\begin{equation}
\label{qdd_Eq7}
\hat{E}= \hat{U}(\pi,0)=\left [{\begin{array}{cc} 
          \\
          0  &  1 \\
        -1  &  0
        \end{array}}
        \right ]
\end{equation}
\begin{equation}
\label{qdd_Eq8}
\hat{A}= \hat{U}(0,\pi/2)=\left [{\begin{array}{cc} 
          \\
          i  &  0 \\
          0  &  -i
        \end{array}}
        \right ]
\end{equation}

\begin{figure}[]
\centering
\includegraphics[scale=.6]{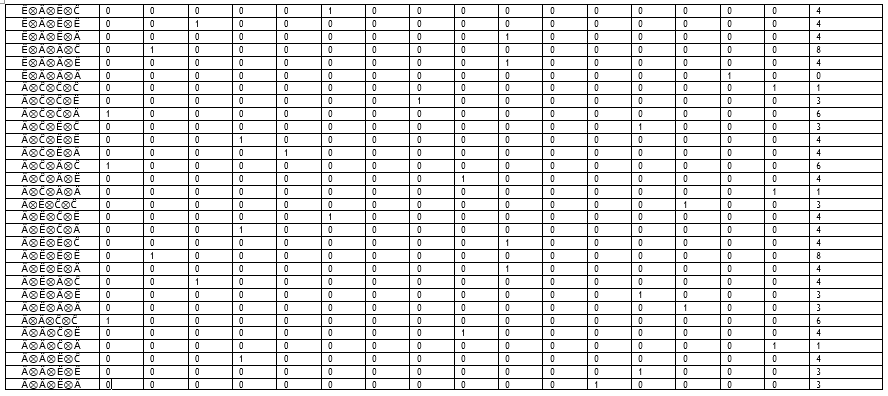}
\caption{Joint probabilities of last 33 strategies of diner's dilemma for n=4 (along with the payoff value of Doug).}
\label{qdd_Fig10}
\end{figure}

$\hat{C}$ and $\hat{E}$ are used to place the order for the cheap and expensive foods respectively. The joint probabilities for the set of different strategies are calculated and shown in Figs. \ref{qdd_Fig9} and \ref{qdd_Fig10}. We get a total of 81 different strategies with which Alice, Bob, Colin and Doug can play. In a list of 81 strategies, we calculate the payoff of Doug.
The payoff table for the quantum model is given in Fig. \ref{qdd_Fig3}. The Payoff is written in order of Alice, Bob, Colin and Doug. In this table, we can observe that there are 8 \textit{Pareto Optimal} points (those are underlined). Let us say Alice, Bob and Colin choose $\hat{E}$, $\hat{E}$ and $\hat{E}$ respectively, then Doug's best reply is $\hat{A}$ and if they choose $\hat{C}$, $\hat{C}$ and $\hat{C}$ then Doug's best reply is $\hat{E}$. If any two of them choose $\hat{C}$ (let us say Alice and Bob), then the best reply for the rest of the two players (Colin and Doug) will be either both $\hat{C}$ or both $\hat{A}$. The two cases can be achieved only by mutual cooperation among the players. Therefore, $\hat{E}\otimes\hat{E}\otimes\hat{E}\otimes\hat{E}$ is no longer a Nash equilibrium point. A new Nash equilibrium point $\hat{A}\otimes\hat{A}\otimes\hat{A}\otimes\hat{A}$ is appeared which gives a payoff value of 6 to all the players.
\begin{equation}
\label{qdd_Eq9}
                    Pf_{i}(\hat{A}\otimes\hat{A}\otimes\hat{A}\otimes\hat{A}) = 6.
\end{equation}
where i is  A,B,C or D.

\begin{equation}
\label{qdd_Eq10}
                     Pf_{A}(\hat{X}\otimes\hat{A}\otimes\hat{A}\otimes\hat{A}) \leq  Pf_{A}(\hat{A}\otimes\hat{A}\otimes\hat{A}\otimes\hat{A}).
\end{equation}
\begin{equation}
\label{qdd_Eq11}
                     Pf_{B}(\hat{A}\otimes\hat{X}\otimes\hat{A}\otimes\hat{A}) \leq  Pf_{B}(\hat{A}\otimes\hat{A}\otimes\hat{A}\otimes\hat{A}).
\end{equation}
\begin{equation}
\label{qdd_Eq12}
                     Pf_{C}(\hat{A}\otimes\hat{A}\otimes\hat{X}\otimes\hat{A}) \leq  Pf_{C}(\hat{A}\otimes\hat{A}\otimes\hat{A}\otimes\hat{A}).
\end{equation}
\begin{equation}
\label{qdd_Eq13}
                     Pf_{D}(\hat{A}\otimes\hat{A}\otimes\hat{A}\otimes\hat{X}) \leq  Pf_{D}(\hat{A}\otimes\hat{A}\otimes\hat{A}\otimes\hat{A}).
\end{equation}
where $\hat{X}\ $ can be $\hat{E}\ $,$\hat{A}\ $ or $\hat{C}\ $.

It can be seen from Fig \ref{qdd_Fig3} that no player can deviate from $\hat{A}\otimes\hat{A}\otimes\hat{A}\otimes\hat{A}$ and increase his or her payoff without decreasing others' payoff. Thus $\hat{A}\otimes\hat{A}\otimes\hat{A}\otimes\hat{A}$ is the best strategy to play with, which is also one of the eight \textit{Paerto Optimal} points. Therefore, we can say that by performing quantum strategies, the dilemma is removed among the players.

\section{Implementation on IBM Computer \label{qdd_Sec4}}
\begin{figure}[]
\centering
\includegraphics[width=12cm]{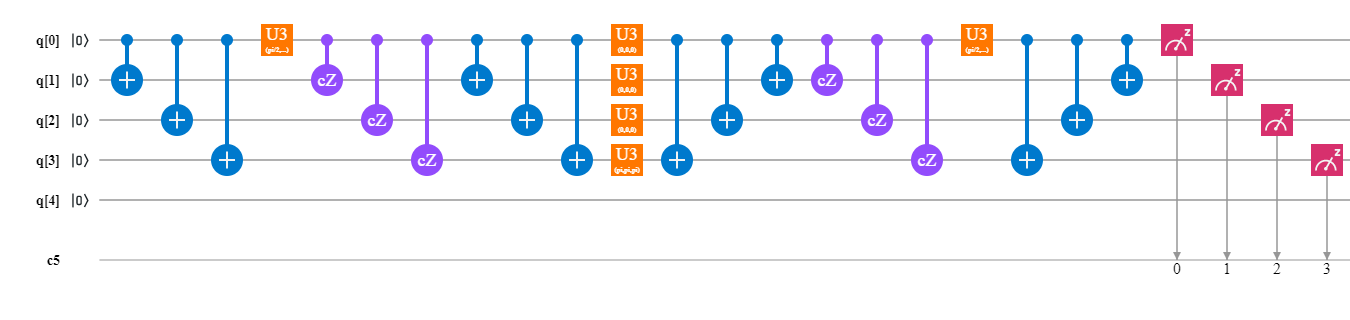}
\caption{Circuit used to implement diner's dilemma on IBM quantum simulator.}
\label{qdd_Fig4}
\end{figure}

\begin{figure}[]
\centering
\includegraphics[width=12cm]{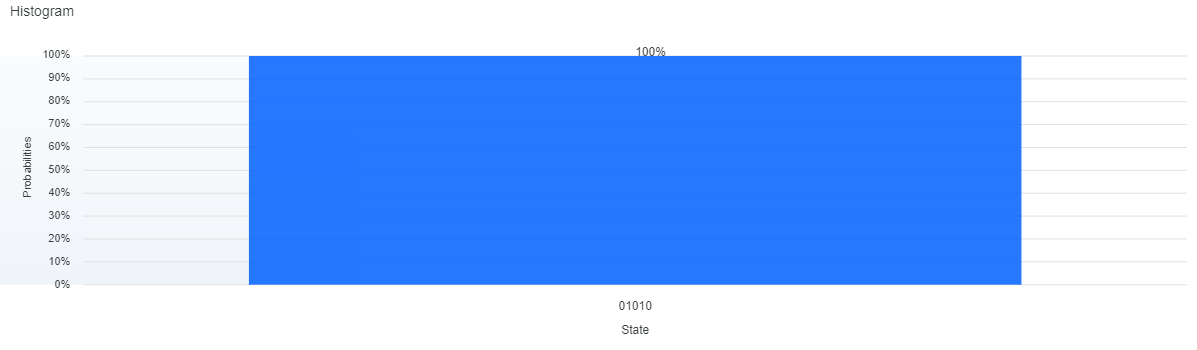}
\caption{Result for the strategy $\hat{C}\otimes\hat{E}\otimes\hat{C}\otimes\hat{E}$.}
\label{qdd_Fig5}
\end{figure}

\begin{figure}[]
\centering
\includegraphics[width=12cm]{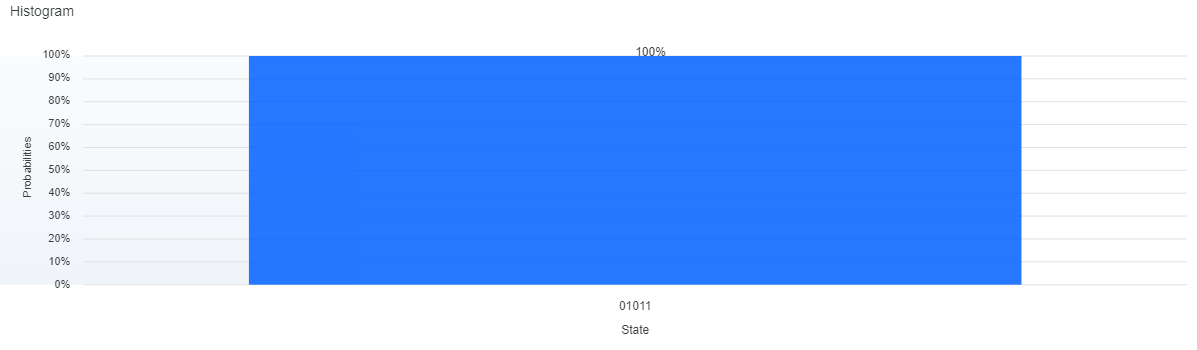}
\caption{Result for the strategy $\hat{C}\otimes\hat{C}\otimes\hat{E}\otimes\hat{A}$.}
\label{qdd_Fig6}
\end{figure}

\begin{figure}[]
\centering
\includegraphics[width=12cm]{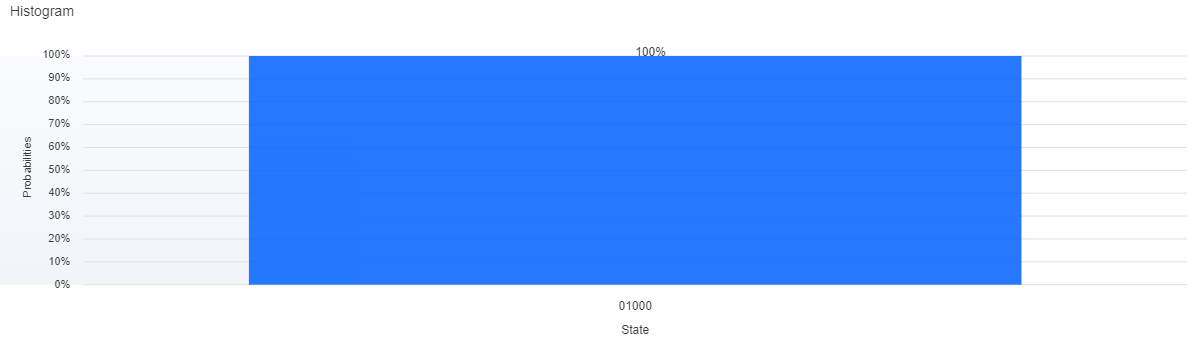}
\caption{Result for the strategy $\hat{C}\otimes\hat{C}\otimes\hat{C}\otimes\hat{E}$.}
\label{qdd_Fig7}
\end{figure}

\begin{figure}[]
\centering
\includegraphics[width=12cm]{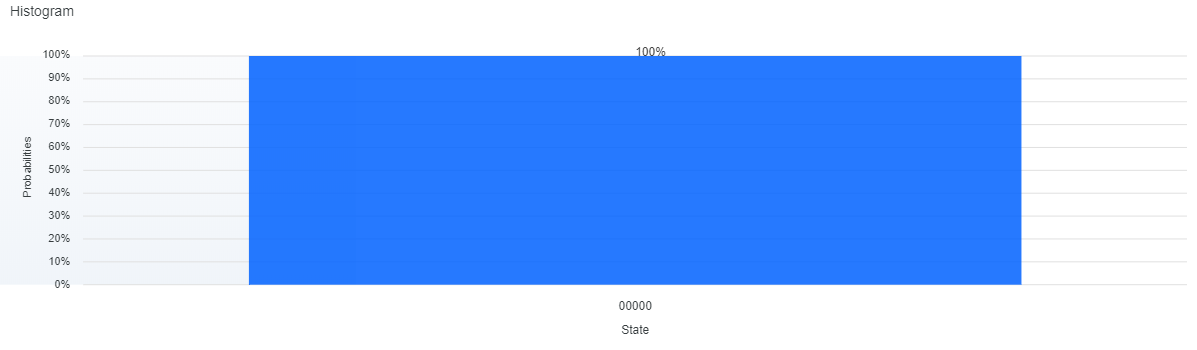}
\caption{Result for strategy $\hat{A}\otimes\hat{A}\otimes\hat{A}\otimes\hat{A}$.}
\label{qdd_Fig8}
\end{figure}

For implementing the above game on the IBM quantum simulator, we use different types of gates (Fig. \ref{qdd_Fig4}) \cite{qdd_chuang},\cite{qdd_khangaming}. For creating an entanglement we use $U_3$ gate with the parameters $(\theta,\phi,\lambda)=(\pi/2,\pi/2,-\pi/2)$, then a series of control-Z gates, and CNOT gates to construct the $\hat{J}$ operator. For different quantum strategy, we use $U_3$ operator with different parameters. For $\hat{C}$, we have $(\theta,\phi,\lambda)$=(0,0,0), for $\hat{E}$ $(\theta,\phi,\lambda)$=$(\pi,\pi,\pi)$ and for $\hat{A}$ $(\theta,\phi,\lambda)=(0,-\pi/2,-\pi/2)$. After then we use $\hat{J}^{\dagger}$ to break the entanglement and finally measure in Z-basis. The circuit is shown in Fig. \ref{qdd_Fig4}. In the circuit, q[0], q[1], q[2] and q[3] belong to Alice, Bob, Colin and Doug respectively. Here, we present the results obtained from the IBM quantum simulator, for four out of the 81 strategies in the form of histograms. The first, second, third and fourth results are of strategies $\hat{C}\otimes\hat{E}\otimes\hat{C}\otimes\hat{E}$ (Fig. \ref{qdd_Fig5}), $\hat{C}\otimes\hat{C}\otimes\hat{E}\otimes\hat{A}$ (Fig. \ref{qdd_Fig6}), $\hat{C}\otimes\hat{C}\otimes\hat{C}\otimes\hat{E}$ (Fig. \ref{qdd_Fig7}) and $\hat{A}\otimes\hat{A}\otimes\hat{A}\otimes\hat{A}$ (Fig. \ref{qdd_Fig8}) respectively.


\begin{figure}[]
\centering
\includegraphics[scale=.6]{qdd_Fig9.png}
\caption{Joint probabilities of first 43 strategy of diner's dilemma for n=4 (along with the payoff value of Doug).}
\label{qdd_Fig9}
\end{figure}
\section{Conclusion \label{qdd_sec5}}
To conclude here, we have demonstrated a quantized version of diner's dilemma problem. It is observed that if the players play this game with the quantum rules, then he or she can escape the dilemma of deciding strategy while ordering food. By applying a quantum strategy, players can reach the \textit{Pareto Optimal} point as well as the Nash equilibrium point. The entanglement of the shared qubits plays an important role in deciding the payoff of the players. The payoff is a function of the extent of entanglement. If entanglement is zero, then the game reduces to the classical scenario and it gives maximum payoff for maximally entangled shared state. We present the circuit implementation of the unitary operators used in the game and design them on the IBM Q simulator. We obtain desired results and verify all the strategies taken by the players. In the present work, we use maximally entanglement state. However, presence of non-maximally entangled states has not been explored till date, which can be done in the future work.

\section*{Acknowledgments}
\label{acknowledgments}
A.A. acknowledges the hospitality provided by IISER Kolkata during the project work. B.K.B. acknowledges the financial support of Institute fellowship provided by IISER Kolkata. We acknowledge IBM Q Experience’s team for providing access to IBM Q quantum simulator and performing the experiments.

\end{document}